\documentclass{emulateapj}
\usepackage{apjfonts}
\usepackage{epsfig}
\usepackage{graphicx}

\newcommand{\cubecm}{\ifmmode{~{\rm cm^{-3}}}\else{~cm$^{-3}$}\fi}
\newcommand{\kms}{\ifmmode{~{\rm km~s^{-1}}}\else{~km s$^{-1}$}\fi}
\newcommand{\lsim}{\lower0.3em\hbox{$\,\buildrel <\over\sim\,$}}
\newcommand{\gsim}{\lower0.3em\hbox{$\,\buildrel >\over\sim\,$}}

\newcommand{\rvir}{\ifmmode{{\rm r_{vir}}}\else{r$_{\rm{vir}}$}\fi}
\newcommand{\mvir}{\ifmmode{{\rm M_{vir}}}\else{M$_{\rm{vir}}$}\fi}
\newcommand{\beq}{\begin{equation}}
\newcommand{\eeq}{\end{equation}}

\shorttitle{FIRST STELLAR MASS BLACK HOLES}
\shortauthors{ALVAREZ, WISE \& ABEL}

\begin{document}

\title{Accretion onto the first stellar mass black holes}

\author{Marcelo A. Alvarez\altaffilmark{1},
John H. Wise\altaffilmark{1,2}, and
  Tom Abel\altaffilmark{1}}

\altaffiltext{1}{Kavli Institute for Particle Astrophysics and
  Cosmology, Stanford University, Menlo Park, CA 94025}
\altaffiltext{2}{Current address: Laboratory for Observational
  Cosmology, NASA Goddard Space Flight Center, Greenbelt, MD 21114}
\email{malvarez@slac.stanford.edu}

\begin{abstract}
The first stars in the universe, forming at redshifts $z>15$ in
minihalos with masses of order $10^{5-6}M_\odot$, may leave behind 
black holes as their remnants.  These objects could conceivably
serve as "seeds" for much larger black holes observed at redshifts 
$z\lesssim 7$. 
We study the growth of the remnant black holes through accretion including
for the first time the emitted accretion radiation with adaptive mesh
refinement cosmological radiation-hydrodynamical simulations.
The effects of photo-ionization and heating dramatically affect the
accretion flow from large scales, resulting in negligible mass growth of the
black hole. We compare cases with the accretion luminosity included and
neglected to show that the accretion radiation drastically changes the
environment within
100 pc of the black hole, where gas temperatures are increased by
an order of magnitude. The gas densities are reduced and
further star formation in the same mini-halo prevented for the two hundred
million years of evolution we followed. These calculations show that even
without the radiative feedback included most seed black holes do not gain
mass as efficiently as has been hoped for in previous theories,
implying that black hole remnants of Pop III stars that formed in
minihalos are not likely to be the origin of miniquasars. Most
importantly, however, these calculations demonstrate that if early stellar mass
black holes are indeed accreting close to the Bondi-Hoyle rate with ten
percent efficiency they have a dramatic local effect in regulating star
formation in the first galaxies. 


\end{abstract}

\keywords{cosmology: theory --- galaxies: formation --- black holes:
  formation}

\section{INTRODUCTION}

\begin{figure*}
  \begin{center}
  \includegraphics[width=.75\textwidth]{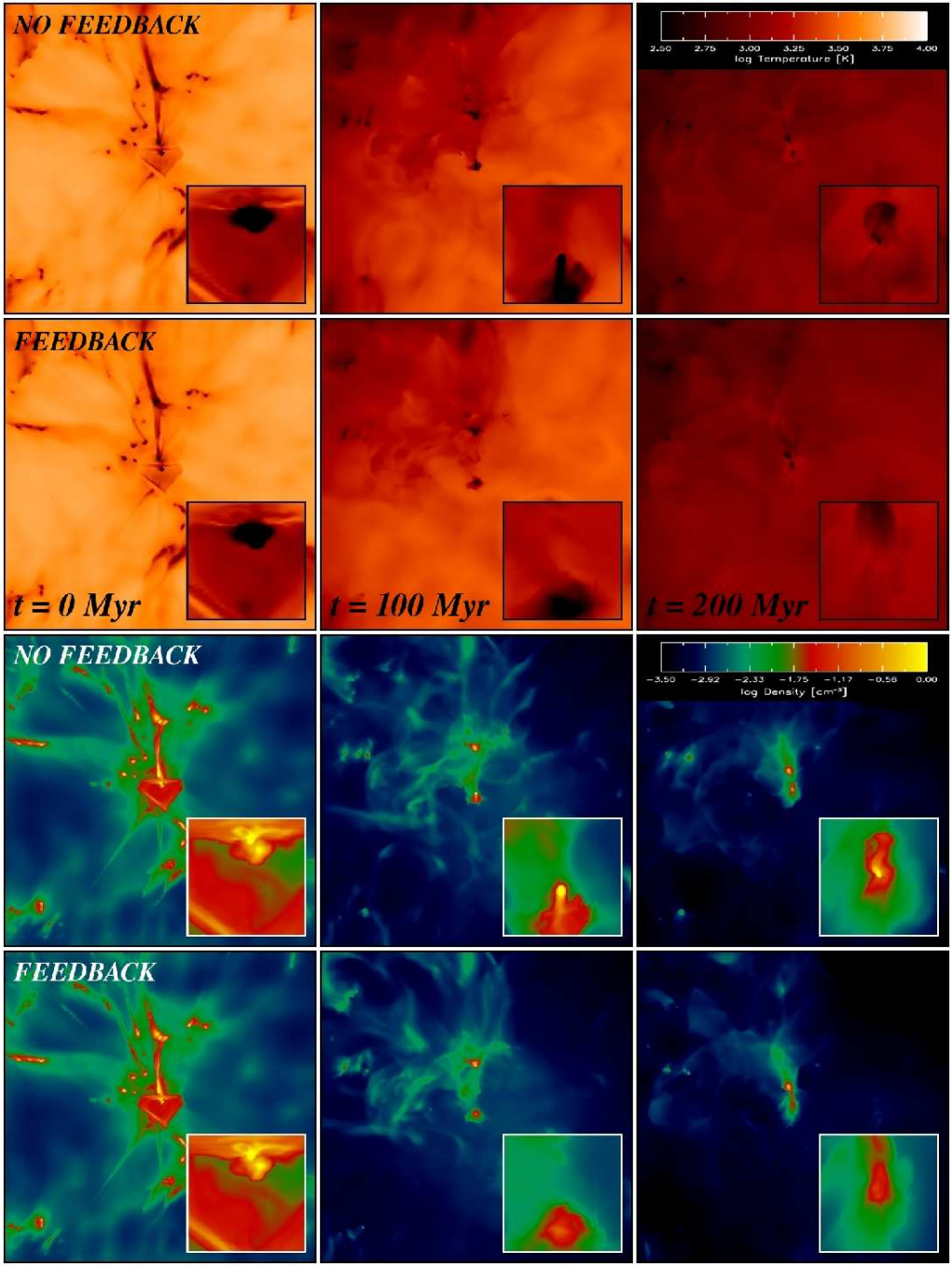}
  \end{center}
  \caption{Density-squared weighted rojections of temperature (top two rows)
    and hydrogen number density (bottom 
    two rows) of cubic regions 87.5 (large panels) and 6.25
    (insets) comoving kpc on a side, centered on the position of the black
    hole at 0 (left -- $z\sim 18$) 100 (center -- $z\sim 14$) and 200
    (right -- $z\sim 11$) Myr after the
    formation of the black hole.  Top rows are without feedback, while
    lower rows are with feedback.  While the large scale structure of
    the volume is hardly affected by the feedback, the insets, on
    scales of approximately 300 proper pc, show a substantial heating and
    density reduction within the host minihalo.}
  \label{panels}
\end{figure*}

The first stars in the universe may have left a population of
black holes as their remnants.  This possibility demands we
understand how radiative feedback would have affected their environment and
regulated accretion.
Luckily, there is a relative amount of theoretical consensus
regarding the nature of first star formation, offering a solid
foundation for further investigation into their black hole remnants.

The first stars probably formed at redshifts $z>15$ in
minihalos with mass $10^{5-6}M_\odot$, and were likely massive, 
with $30 \lesssim M_*/M_\odot \lesssim 300$
\citep{abel/etal:02,bromm/etal:02}. They influenced
subsequent structure formation through their UV radiation 
heating the surrounding gas and initiating reionization
\citep{whalen/etal:04,kitayama/etal:04,alvarez/etal:06,
johnson/bromm:07,abel/etal:07,yoshida/etal:07,wise/abel:07,wise/abel:08}.
If these population III (Pop III) stars were less massive than 140~
$M_\odot$ or more massive than 260~$M_\odot$, they likely would
collapse to form a black hole at the end of their lifetime
\citep{heger/etal:03}.

The presence of these black holes could have had dramatic consequences.
If a 100-$M_\odot$ black hole is radiating at the Eddington limit with
ten per cent radiative efficiency, beginning at $z>20$, it will
attain a final mass at 
$z=6.4$ of greater than $10^9 M_\odot$, implying that accretion onto
these black holes is a viable explanation for the $z\gsim 6$ quasars
\citep[e.g.,][]{haiman/loeb:01,volonteri/rees:06}.  Such
efficient accretion would mean that ``miniquasars'' may have
been abundant during reionization. 


Feedback from black hole accretion is a very active field, 
focusing mainly on its relationship to galaxy formation
\citep[e.g.,][]{silk/rees:98, ciotti/ostriker:01}. 
A typical approach is that described by \citet{springel/etal:05}, where
some fraction of the rest-mass energy accreted onto the black hole is
deposited as thermal energy within a finite volume of surrounding
matter as an additional term in the energy equation.  This technique
was recently used by \citet{li/etal:07} and \citet{pelupessy/etal:07}
to study the growth of Pop III remnant black holes.  While giving
insight into the {\em late time} growth of the seed black holes, these
simulations by design do not address the earliest phases of accretion.

\begin{figure}
  \begin{center}
  \includegraphics[height=.26\textheight]{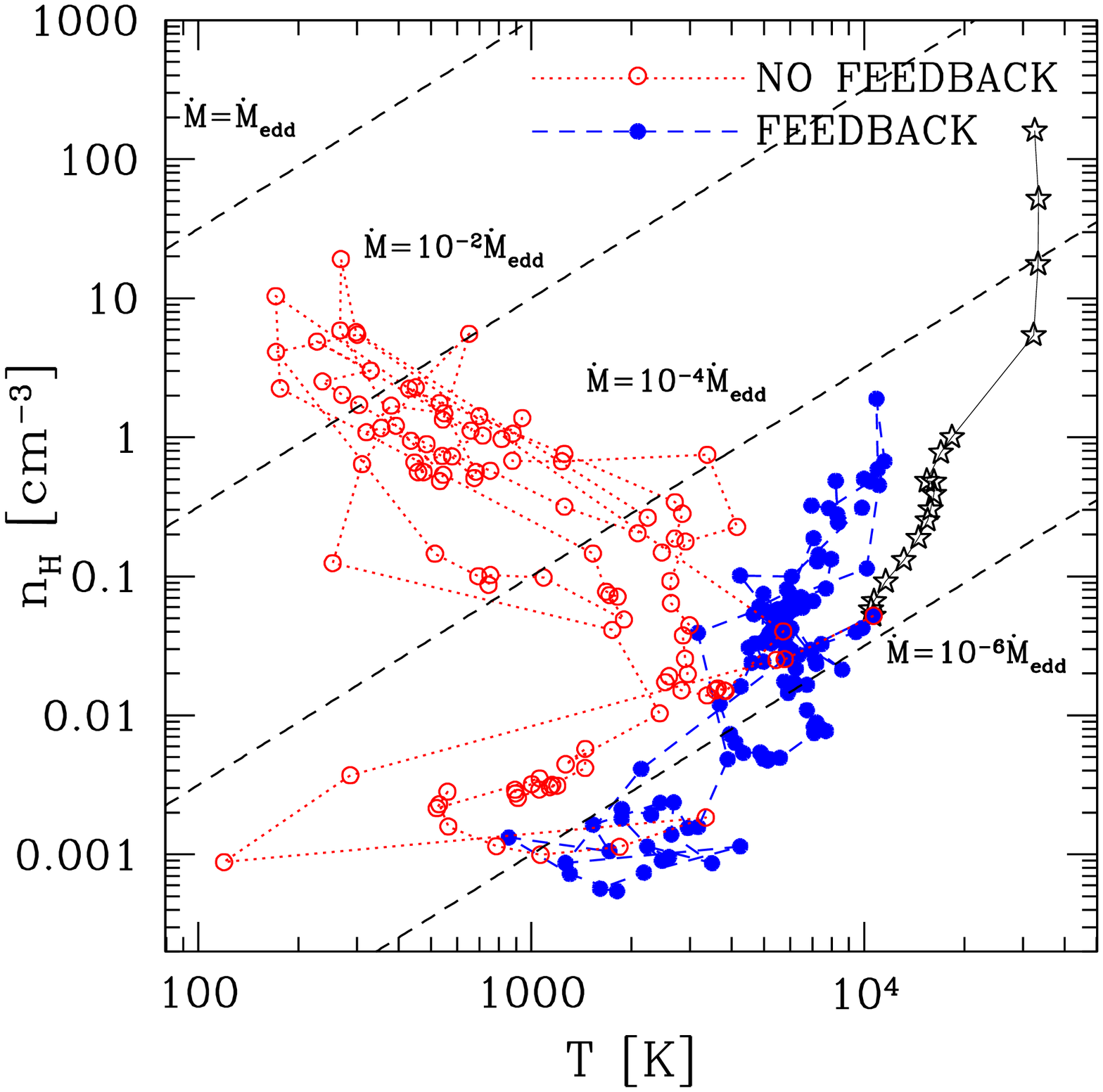}
  \end{center}
  \caption{Density and temperature at position of black
  hole along its path (starts at upper right) with (solid) and
  without (dotted) feedback.  The points along the solid line
  beginning at the top right of the
diagram are during the progenitor
star's lifetime, and each point thereafter is spaced approximately 2
Myr apart.  Also shown as dashed
  lines are loci of fixed Bondi accretion rate, in units of
  the mass accretion rate if the black hole were shining at the
  Eddington luminosity with radiative efficiency, $\epsilon=0.1$.
}
  \label{phase}
\end{figure}


In this {\em Letter} we focus on the evolution of
remnant black holes immediately following the death
of their progenitor Pop III stars.
Our simulations 
account for the ionizing radiation from the Pop III star itself, which
photoevaporates the gas within its host halo.  Because the simulations
are run from cosmological initial conditions, we can follow the fate
of the black hole as it encounters cold, dense gas from nearby halos,
as well as the recombining gas in the relic H~II
region.  We model the ionizing radiation from
accretion onto the black hole by full radiative transfer coupled to
the hydrodynamics, revealing how it regulates black hole growth
and alters the nearby environment.  We present our
simulation method in \S 2 and our results in \S 3.  We conclude with a
discussion in \S 4.  We will present the results of two otherwise identical
simulations runs, one in which ionizing radiation from the accretion
is included (``feedback''), and one in which it is not (``no
feedback'').  The simulations with no feedback were first presented in
\citet{alvarez/etal:08}.

\section{Method}

We use the code Enzo\footnote{http://lca.ucsd.edu/projects/enzo}
\citep{oshea/etal:04} to simulate the formation of the first star
within a periodic volume 250 comoving kpc on a side.  The dark matter
particle mass within the central $\sim 90$ comoving kpc, where
refinement is allowed to proceed adaptively, is $30 M_\odot$.  A
fully-formed 100 $M_\odot$-star is assumed to form when the
overdensity is $5\times 10^5$ and the molecular hydrogen mass fraction
is greater than $5\times 10^{-4}$.  The star is assigned an ionizing
photon luminosity of $1.2\times 10^{50}$~s$^{-1}$ and shines for its
main sequence lifetime of 2.7~Myr. Adaptive
ray-tracing is used for the radiative transfer of ionizing radiation, 
which is coupled to the hydrodynamics via the energy
equation and integrated into Enzo's primordial chemistry solver.
For more
details on the simulation set up and method, see
\cite{abel/etal:07} (the set up is just the same as there, but
with a different set of random modes for the initial conditions).
After the lifetime of the star, radiation is turned off and
it is assumed to collapse directly to a black hole with the same
mass, position, and velocity as the progenitor. We track the position
of the black hole as the relic HII region evolves for 210 Myr from
$z\sim 18$ to $z\sim 11$.

We base our model for the spectrum of
accreting gas on an approximate form of the multi-color disk +
power-law model adopted by \cite{kuhlen/madau:05}.  We
take $L_\nu\propto \nu$ for $h\nu < 200$~eV, $L_\nu\propto \nu^{-1}$ for
200 eV $< h\nu < 10$~keV, and $L_\nu = 0$ for $h\nu > 10$~keV.  This
implies a mean photoelectron energy of 460 eV, which we take as our
monochramatic photon energy.  While admittedly crude, our
model is adequate in this context,
where we focus on the larger scale environment on 10-1000 pc
scales. Neglecting dissociating radiation places an upper limit
on positive feedback effects, and is therefore conservative in this
respect.  The normalization of the total luminosity is obtained by assuming that
ten per cent of the rest mass energy of matter falling in is converted
to radiation, and the accretion rate is determined from the
Bondi-Hoyle \citep{bondi/hoyle:44,bondi:52} formula,
$\dot{M}_{BH}=4\pi G^2M^2\rho/(c_s^2+v_{\rm rel}^2)^{3/2}$, where
$c_s$ is the sound speed of the gas and $v_{\rm rel}$ is the velocity of the
black hole relative to nearby gas.  The Bondi radius, $r_{\rm B}\simeq 
10^{17}$~cm~$(M_{\bullet}/100 M_\odot)(1000~{\rm K}/T)$, is close to our
resolution limit, indicating the local conditions at the black hole
are appropriate for determining the accretion rate.
Snapshots of the simulation are
shown in Fig. \ref{panels}.

\section{Results}

The black hole is tightly bound to the center of the dark matter 
halo in which its progenitor formed for the duration of the simulation.
Its path through density-temperature space is shown in
Fig. \ref{phase}, in cases with and without feedback from
accretion included. Also plotted are lines of constant 
Bondi accretion rate in units of the
Eddington accretion rate, $\dot{M}_{\rm
  Edd}=4\pi GMm_p(1-\epsilon)/(\epsilon\sigma_T c)$, where the
radiative efficiency is set to $\epsilon=0.1$.  For $M=100$~$M_\odot$,
the Eddington accretion rate is $\simeq 2\times 10^{-6}
M_\odot$~yr$^{-1}$.  

\begin{figure}
  \begin{center}
  \includegraphics[height=0.45\textheight]{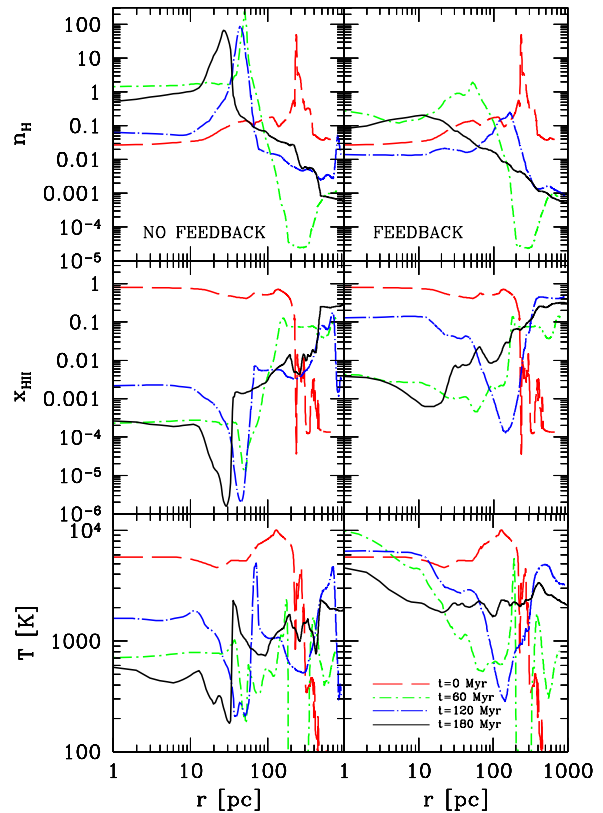}
  \end{center}
  \caption{Profiles of hydrogen number density (top), ionized fraction (middle),
  and temperature (bottom), along line
  connecting the black hole (at $r=0$) to the highest density point
  nearby, with (right) and without (left) feedback.  The linestyle
  distinguish times after the black hole is formed.  Their times are
  given in the lower right panel.}  
\label{profs}
\end{figure}

Initially,
the black hole is bathed in hot, ionized gas, with a temperature $\sim
10^4$~K and hydrogen number density $\sim 1$~cm$^{-3}$.  The
photoevaporative wind from the first star continues outward under its
own inertia with a velocity $\sim 10$~km~s$^{-1}$.
The black hole eventually encounters
a nearby halo with which its own host halo is merging, after which the
evolution depends on whether or not radiative
feedback due to accretion is included.  Without feedback included,
the path
through density-temperature space oscillates between cold dense gas
close to the neighboring halo's center, to the warmer more rarefied
gas in the ambient relic H~II region, due to the orbital motion of the
merging halos.  With feedback included, however, the black hole stays
in relatively low-density, high-temperature gas for the entire time,
partially ionizing and heating gas within hundreds of pc
(Figs. \ref{panels} and \ref{profs}).



\begin{figure}[t]
  \begin{center}
  \includegraphics[height=0.34\textheight]{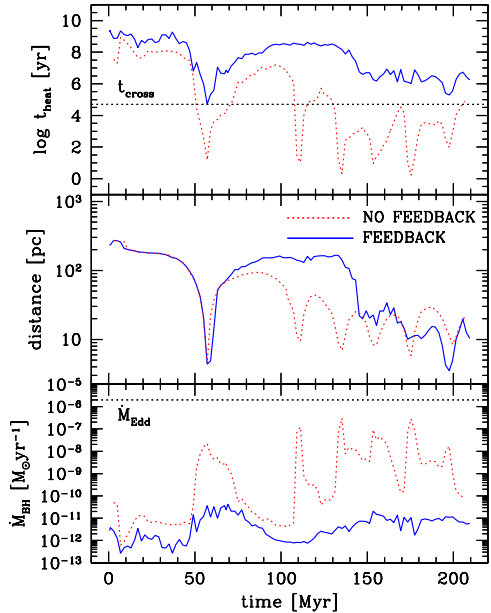}
  \end{center}
  \caption{Evolution of heating time (top), distance from black hole
    to clump (middle), and accretion rate (bottom).  Solid lines are
  with feedback, dotted lines are without feedback. The horizontal
line in the top panel is the sound crossing time across a 1 pc
photoionized clump, while the one in the lower panel is the Eddington 
accretion rate for a 100 $M_\odot$ black hole.}
\label{time}
\end{figure}

Shown in Fig. \ref{time} are distance to the nearest high density peak
(``clump''), the heating time at the clump, 
and accretion rate (in units of $\dot{M}_{\rm Edd}\simeq
2\times 10^{-6}~M_\odot$yr$^{-1}$) .  This figure illlustrates the
in-spiralling motion of the clump and black hole as their host halos
merge.  When the clump approaches, the accretion rate rises,
accompanied by a shorter heating timescale.  The dotted line in the top
panel is the sound crossing time of a 1 pc clump at $\sim$30,000 K.  Before
the clump approaches the black hole for even the first time, the
heating timescale becomes shorter than its sound crossing time, and
photoevaporation ensues.  For a radiative
efficiency of ten percent, self-regultation of accretion onto black
holes within minihalos occurs at accretion rates some five orders of
magnitude below the Eddington rate, at about $10^{-11} M_\odot{\rm yr}^{-1}$.

\section{Discussion}

We have presented results from a simulation of radiative feedback from
accretion onto a remnant black hole of an early Pop III star. This feedback 
makes it even more
difficult for the black hole to encounter cold, dense gas, further
depleting the reservoir of gas available for accretion.  Including
radiative feedback lowers peak accretion 
rates from $\sim 10^{-7} M_\odot{\rm yr}^{-1}$ to $\sim 10^{-11}
M_\odot{\rm yr}^{-1}$, indicating that radiative feedback is very
efficient at halting accretion.  Furthermore, the feedback is sufficiently
strong to prevent further cooling and star formation within its
host minihalo, inhibiting star formation in the halo over the 200
million years we have evolved it.  
Further star formation can only occur in larger systems, for
which the dark matter potential well is deep enough or within which
star formation can occur outside of the center of the halo.  

These results are in agreeement with previous work, which indicated
that massive Pop III stars forming in
minihalos with mass $\sim 10^6 M_\odot$ photoevaporate the halo gas
from the inside out, leaving any remnant black hole starved, at least
momentarily, of accreting material
\citep{whalen/etal:04,kitayama/etal:04,oshea/etal:05}.
In particular, \citet{johnson/bromm:07} showed that this
photevaporation was likely to delay significant accretion for on the
order of 100 million years, creating an early ``bottleneck'' in the
black hole's growth.  Our results imply that the situation is
likely to be even more drastic, because radiative feedback is so
effective at preventing gas from reaching the center of the halo, even
without taking into account the effect of radiation pressure
\citep[e.g.,][]{haehnelt:95,dijkstra/loeb:08,milos/etal:08}.  An appealing alternative
scenario for seed black hole formation and accretion was recently
discussed by \citet{dijkstra/etal:08}, in which intermediate mass black
holes can form by direct collapse in halos
\citep[e.g.,][]{bromm/loeb:03}.  More detailed numerical
simulations of the kind presented here will be required to test
this hypothesis.


The possibility that accretion onto these black holes could power
miniquasars has 
received a great deal of attention lately, since their radiation 
could have significantly increased the ionizing photon budget of the
universe and extended the reionization process
\citep{ricotti/ostriker:04,madau/etal:04}.  
Because their radiation would have a long
mean free path, they would also change the topology of reionization, 
blurring the distinction between ionized and neutral gas.
They could also change the pattern of fluctuations in the
21-cm background before reioinization, as pointed out by
\citet{chuzhoy/etal:06} and \citet{chen/miralda-escude:06}.  
Black hole remnants of Pop III stars that formed in
minihalos are not likely to be the origin of the miniquasars.
As we have shown, significant accretion onto these black holes will be
delayed until they fall into more massive dwarf galaxies, which are
likely to be the sites of significant star formation, perhaps of Pop
II stars.  Black hole remnants from these objects are likely to be
both more numerous, and in locations of higher gas density than the
black holes that formed in minihalos, thus replacing the Pop III seeds
in models like that of \citet{tanaka/haiman:08}.

In future
work it will be necessary to determine at which values of the radiative 
efficiency feedback from black hole accretion becomes  
irrelevant. More importantly, we must better constrain 
the angular distribution (i.e. beaming), intensity, and spectral shape of the
radiation produced by the accretion flow around massive early stellar
mass black holes in the 
environments we detailed here. Such modeling will require far more  
spatial and temporal dynamic range than has hitherto been feasible.

\vspace{-0.5cm}
\acknowledgments{This work was supported by NSF CAREER award
  AST-0239709 from the National Science Foundation.  M.A.A. thanks
  Zoltan Haiman, Mike Kuhlen, Avi Loeb, Piero Madau, Peng Oh, Jerry 
Ostriker, and Naoki Yoshida for helpful comments on an earlier draft.  
We are grateful
  for the continuous support from the computational team at SLAC.  We
  performed these calculations on 16 processors of a SGI Altix 3700
  Bx2 at KIPAC at Stanford University.}

{} 

\end{document}